\documentclass[twocolumn,showpacs,aps,prb,reprint,amsfonts,amsmath,amssymb,floatfix,groupedaddress, longbibliography]{revtex4-2} % arxiv luatexを抜く
\usepackage{siunitx}
\usepackage{color}
\usepackage{hhline}
\usepackage{mathrsfs}
\usepackage{graphicx} % for hyperref
\usepackage{adjustbox}
\usepackage{dcolumn}
\usepackage{bm}
\usepackage{multirow}
\usepackage{booktabs}
\usepackage{afterpage}
\usepackage{amsmath}
\usepackage{ulem}
\usepackage{physics}
\usepackage{here} 
\usepackage[inline]{asymptote}
\usepackage{caption}
\usepackage[subrefformat=parens]{subcaption}
\usepackage[version=4]{mhchem}

\graphicspath{{figure/}{table/}}
\usepackage{multirow}
\usepackage[subpreambles=true,sort=true]{standalone}

\usepackage{caption} 
\usepackage{comment}
\usepackage[colorlinks=true,citecolor=blue,linkcolor=blue,urlcolor=blue]{hyperref}
\usepackage{cleveref}
\crefname{equation}{Eq.}{Eq.}
\crefname{figure}{Fig.}{Fig.}
\crefname{table}{Table}{Table}
\crefname{section}{Sec.}{Sec.}

\Crefname{equation}{Equation}{Equation}
\Crefname{figure}{Figure}{Figure}
\Crefname{table}{Table}{Table}
\Crefname{section}{Section}{Section}

\usepackage{threeparttable}

\begin{document}
\title{Lattice dielectric properties of rutile \ce{TiO2}: \\ 
First-principles anharmonic self-consistent phonon study}

\author{Tomohito Amano$^{1}$ }
\author{Tamio Yamazaki$^{2}$ }
\author{Ryosuke Akashi$^{1,3}$ }
\author{Terumasa Tadano$^{4}$ }
\author{Shinji Tsuneyuki$^{1}$ }
\affiliation{$^1$Department of Physics, The University of Tokyo, Hongo, Bunkyo-ku, Tokyo 113-0033, Japan}
\affiliation{$^2$JSR Corporation, RD Technology and Digital Transformation Center 1-9-2, Higashi-Shinbashi, Minato-ku, Tokyo 105-8640 Japan }
\affiliation{$^3$Quantum Materials and Applications Research Center, National Institutes for Quantum Science and Technology, 2-10, Ookayama, Meguro-ku, 152-0033, Tokyo, Japan}
\affiliation{$^4$Research Center for Magnetic and Spintronic Materials, National Institute for Materials Science, Tsukuba 305-0047, Japan}

\date{\today}
\onecolumngrid

\begin{abstract}
We calculate the lattice dielectric function of strongly anharmonic rutile \ce{TiO2} from \textit{ab initio} anharmonic lattice dynamics methods. Since an accurate calculation of the $\Gamma$ point phonons is essential for determining optical properties, we employ the modified self-consistent approach, including third-order anharmonicity as well as fourth-order anharmonicity. The resulting optical phonon frequencies and linewidths at the $\Gamma$ point much better agree with experimental measurements than those from a perturbative approach.  We show that the four-phonon scattering process contributes as much as the third-order anharmonic term to phonon linewidths. Furthermore, incorporating the frequency dependence of phonon linewidth reveals that experimentally known but unidentified peaks of the dielectric function are due to two-phonon process. This work emphasizes the importance of a self-consistent approach in predict the optical properties of highly anharmonic materials.
\end{abstract}

\twocolumngrid
 
\maketitle

\section{Introduction}\label{sec:intro}

Titanium dioxide (\ce{TiO2}) is a polar semiconductor, which has been studied extensively from both experimental and theoretical perspectives for its phenomenal dielectric constants of $111$ and $250$ along the $x$ and $z$ axes, respectively. The consequent high refractive index is advantageous for various technological applications such as pigments and capacitors. Rutile \ce{TiO2} thin film has also attracted attention as a high-$\kappa$ dielectric material for DRAM~\cite{kim2004High}.

\begin{figure}[t]
\includegraphics[]{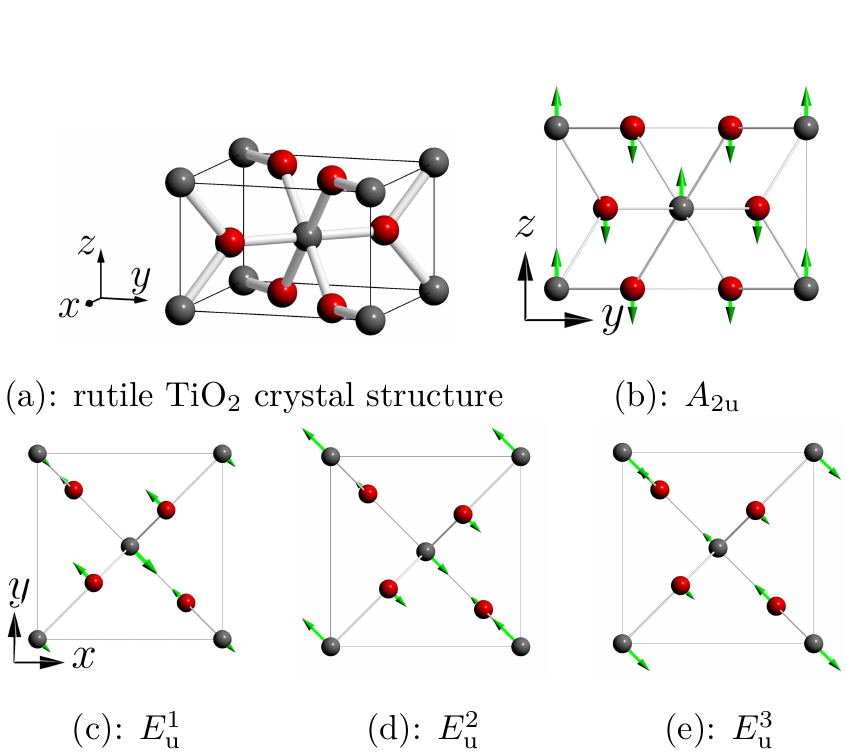}
% \hfill
% \centering
% \begin{subcaptionblock}{0.6\linewidth}
% \centering
% \includegraphics[]{rutile_axis-1_0.pdf}
% \subcaption{}
% \end{subcaptionblock}\hfill
% \begin{subcaptionblock}{0.4\linewidth}
% \centering
% \includegraphics[width=\linewidth]{A2u.pdf}
% \subcaption{$A_{\mathrm{2u}}$}
% \end{subcaptionblock}\hfill
% \begin{subcaptionblock}{0.3\linewidth}
% \centering
% \includegraphics[width=\linewidth]{E1u.pdf}
% \subcaption{$E^{1}_{\mathrm{u}}$}
% \end{subcaptionblock}\hfill
% \begin{subcaptionblock}{0.3\linewidth}
% \centering
% \includegraphics[width=\linewidth]{E2u.pdf}
% \subcaption{$E^{2}_{\mathrm{u}}$}
% \end{subcaptionblock}\hfill
% \begin{subcaptionblock}{0.3\linewidth}
% \centering
% \includegraphics[width=\linewidth]{E3u.pdf}
% \subcaption{$E^{3}_{\mathrm{u}}$}
% \end{subcaptionblock}\hfill
\caption{(a) The unit cell of rutile \ce{TiO2}, which contains two titanium atoms (black) and four oxygen atoms (red). (b-e) Schematic views of atomic displacements for the $A_{\mathrm{2u}}$ mode and the three $E_{\mathrm{u}}$ phonon modes. }
\label{fig:crystal}
\end{figure} 

The importance of rutile \ce{TiO2} has instigated several experimental and theoretical studies on dielectric properties~\cite{devore1951Refractive, parker1961Static, spitzer1962Far, barker1963Far, samara1973Pressure, gervais1974Anharmonicity,gervais1974Temperature, matsumoto2008Analysis, schoche2013Infrared, kanehara2015Terahertz}. The large dielectric constant directly links to substantial Born effective charges and a low-frequency transverse optical phonon mode ($A_{\mathrm{2u}}$, see \cref{fig:crystal}). The frequency of the $A_{\mathrm{2u}}$ phonon rapidly increases with increasing temperature~\cite{traylor1971Lattice}, as in the case of ferroelectric crystals, and is accompanied by a decrease in the static dielectric constant. However, unlike ferroelectric crystals, the frequency of the $A_{\mathrm{2u}}$ phonon does not become zero with lowering temperature, and therefore the system does not undergo a phase transition. Several perovskites (e.g., \ce{KTaO3}) are known as such materials and are called incipient ferroelectric. The strong anharmonicity of the lattice~\cite{samara1973Pressure} is the reason for such remarkable temperature-dependent behavior. Gervais and Piriou~\cite{gervais1974Anharmonicity, gervais1974Temperature} applied the four-parameter semi-quantum model (FPSQ) as a model of the dielectric function and successfully fitted experimental reflectivity data. The model partially accounts for anharmonic effects employing different damping parameters for each transverse optical (TO) and longitudinal optical (LO) phonon. The FPSQ model studies~\cite{gervais1974Anharmonicity, gervais1974Temperature, matsumoto2008Analysis, schoche2013Infrared} showed a marked difference in damping parameters between each LO and TO phonon, indicating that the conventional harmonic vibration model breaks down, especially for the $A_{\mathrm{2u}}$ phonon mode. 

The first \textit{ab initio} study on the lattice dynamics of rutile \ce{TiO2} by Lee et al.~\cite{lee1994Dielectric} successfully calculated large Born effective charges and static dielectric constant, which led to many other studies on harmonic phonon properties of rutile \ce{TiO2} using input from first-principles calculations~\cite{montanari2002Lattice, montanari2004Pressureinduced, sikora2005Initio, mitev2010Soft, lee2011Influence, grunebohm2011Firstprinciples, wehinger2016Soft, zhang2019Subtlety}. These calculations unveiled the importance of the mixed covalent and ionic bonding of $s$ orbitals of oxygen and $d$ orbitals of titanium, the cause of which is large polarizability due to long-range Coulomb interactions between the ions. The high Born effective charges could be caused by the dynamical transfer of electrons associated with atomic displacements. Therefore, careful convergence testing is required to get meaningful results. Also, the phonon frequencies of the $A_{\mathrm{2u}}$ and TA phonons show strong strain dependencies~\cite{montanari2004Pressureinduced,mitev2010Soft,wehinger2016Soft}. For example, the generalized gradient approximation (GGA) of Perdew-Burke-Ernzerhof yields an overestimation of the lattice constants, resulting in the $A_{\mathrm{2u}}$ phonon with imaginary frequency~\cite{montanari2002Lattice}. These results indicate that the phonon frequencies are sensitive to exchange-correlation functionals and the accuracy of the pseudopotentials~\cite{lee2011Influence}. While LDA functionals are often used in previous calculations and have been successful in describing lattice dynamics despite the underestimation of the lattice constants, recent works~\cite{zhang2019Subtlety,lee2011Influence} revealed that meta-GGA and hybrid functionals give us more accurate lattice constants.

Recently, an \textit{ab initio} computational framework of phonon anharmonicity has been developed to calculate lattice thermal conductivity, phonon lifetime, and other phonon-related properties. In the framework, harmonic and anharmonic interatomic force constants (IFCs) are extracted from first-principles density functional theory (DFT) or density functional perturbation theory (DFPT) calculations. Computing a dynamical matrix from harmonic IFCs give us frequencies and eigenvectors of ordinary harmonic phonons, whereas anharmonic IFCs determine self-energies that cause the frequency shifts and linewidths.

Regarding rutile \ce{TiO2}, several previous studies~\cite{torres2019Thermala, fu2022Finitetemperature} have calculated thermal conductivity using this framework. Fu et al.~\cite{fu2022Finitetemperature} found that the finite-temperature effective IFCs~\cite{hellman2011Lattice}, including higher order anharmonicity, are essential for predicting thermal conductivity, whereas calculations only including third-order anharmonicity underestimated the thermal conductivity. This result suggests that including higher-order IFCs explains the lattice properties of rutile \ce{TiO2}. The validity of perturbative approaches taken in previous studies is questionable in highly anharmonic cases such as rutile \ce{TiO2}, where the anharmonic term contributes as much as $20\%$ of the $A_{\mathrm{2u}}$ phonon frequency at room temperature. The self-consistent phonon (SCPH) theory~\cite{tadano2015Selfconsistent,tadano2018FirstPrinciplesa}, which includes the frequency shift associated with fourth-order anharmonicity in a self-consistent manner, can treat such strongly anharmonic crystals. Recently, the SCPH+B theory has been developed, including the frequency shift associated with third-order anharmonicity within a quasiparticle approximation~\cite{tadano2022FirstPrinciples}. It could describe the possible cancellation of frequency shifts of third and fourth-order anharmonicity in the $A_{\mathrm{2u}}$ phonon.

While the accurate SCPH theory has been successful in many thermal conductivity calculations, there have been few such attempts for lattice dielectric properties~\cite{vanroekeghem2020Highthroughput,choi2021Strain}, only discussing static dielectric constants. Perturbative approaches have been applied to lattice dielectric properties of weakly anharmonic materials. The Lorentz oscillator or FPSQ model studies revealed that the anharmonic term of four phonon scattering (4ph) is not negligible for calculating optical properties~\cite{yang2020Observationa, tong2020Firstprinciples}. Fugallo et al.~\cite{fugallo2018Infrared} used the Cowley formula~\cite{cowley1963Lattice} to incorporate the frequency dependence of a damping parameter, and successfully obtained the dielectric spectra of \ce{MgO} in good agreement with experiment. Here we aim to study the lattice dielectric properties of strongly anharmonic rutile \ce{TiO2}, where such a perturbative approach does not apply.

In this work, we perform first-principles lattice dynamics calculations to predict the IR optical properties of strongly anharmonic rutile \ce{TiO2}. The second-, third-, and fourth-order IFCs are computed using the least absolute shrinkage and selection operator (LASSO) technique based on first-principles calculations. Phonon frequency shifts and linewidths were calculated using SCPH+B, and the Cowley formula was utilized to calculate the dielectric function. We found that the results of the $\mathrm{r}^2$SCAN functional are in good agreement with experimental measurements and that a self-consistent method is essential for describing the strong anharmonicity of the rutile \ce{TiO2}.
 
\section{Theory}
\subsection{Dielectric properties}
The lattice dielectric function at photon energy $\hbar\omega$ is generally described by the classical Lorentz model 
\begin{align} 
\epsilon(\omega)=\epsilon^{\infty}+\sum_{j}\frac{\Delta\epsilon_j\omega^2_{\vb*{0}j}}{\omega^2-\omega^2_{\vb*{0}j}+i\omega\gamma_{\vb*{0}j}},
\end{align}
where $\omega_{\vb*{q}j}$, $\Delta\epsilon_j$, and $\gamma_{\vb*{q}j}$ are the resonant frequency, the oscillator strength, and the damping (FWHM) of the phonon with wave vector $\vb*{q}$ and mode $j$. $\epsilon^{\infty}$ is the electronic dielectric constant. Although this model can describe dielectric properties qualitatively, it may not work well quantitatively because it is based on the Newton's equation of motion, ignoring the frequency dependence of damping constants.

According to the Maxwell's equations, the poles of a dielectric function are TO phonon frequencies, and the poles of a extinction coefficient $\eta=1/\epsilon$ are LO phonon frequencies. The following factorized form was devised to analyze LO and TO phonons having different phonon frequencies and dampings. This model is called FPSQ, as there are four parameters per mode.
\begin{align}
 \epsilon(\omega)=\epsilon^{\infty}\prod_{j}\frac{\omega_{\vb*{0}j,\mathrm{LO}}^2-\omega^2+i\omega\gamma_{\vb*{0}j,\mathrm{LO}}}{\omega_{\vb*{0}j,\mathrm{TO}}^2-\omega^2+i\omega\gamma_{\vb*{0}j,\mathrm{TO}}}.
\end{align}
When LO-TO splitting is large, namely, $\omega_{\mathrm{LO}} \gg \omega_{\mathrm{TO}}$, the difference in a damping is more pronounced, and the FPSQ model is more suitable than the Lorentz model. Even though the model does not consider the frequency dependence of damping, it successfully explains the experimental values well for a wide range of materials.

On the other hand, Cowley~\cite{cowley1963Lattice} derived an equation incorporating the full frequency dependence of dampings using the anharmonic lattice dynamics theory~\cite{born1955Dynamical} and the linear response theory (see \cref{appendix:A} for derivation),
\begin{align}
 \epsilon_{\alpha\beta}(\omega)=\epsilon^{\infty}_{\alpha\beta}+\frac{1}{v_0}\sum_{j}\frac{S^{j}_{\alpha\beta}}{\left(\omega_{\vb*{0}j}\right)^2-\omega^2-2\omega_{\vb*{0}j} \Sigma_{\vb*{0}j}(\omega)},\label{143252_19Jul22}
\end{align}
where $v_0$ is the volume of the unitcell, $\alpha$ and $\beta$ are Cartesian indices, and $\Sigma(\omega)=-\Delta\omega(\omega) + i\Gamma(\omega)$ is phonon self-energy, where $\Delta\omega(\omega)$ and $\Gamma(\omega)$ are called frequency shift and linewidth, respectively. Phonon lifetime $\tau_{\vb*{q}j}$ is related to linewidth as $\tau_{\vb*{q}j}=1/2\Gamma(\omega_{\vb*{q}j})$, and a damping parameter in the Lorentz model or the FPSQ model holds $\gamma_{\vb*{0}j}=2\Gamma(\omega_{\vb*{0}j})$. Summations are taken only for TO phonons at the $\Gamma$ point. $S$ is called mode-oscillator strength defined as follows~\cite{gonze1997Dynamical}, 
\begin{equation}
 S^{j}_{\alpha\beta}= \left(\sum_{\kappa\alpha'}Z^{*}_{\kappa,\alpha\alpha'}\frac{e_{\kappa\alpha'}(\vb*{0}j)}{\sqrt{m_{\kappa}}}\right)\left(\sum_{\kappa\beta'}Z^{*}_{\kappa,\beta\beta'}\frac{e_{\kappa\beta'}(\vb*{0}j)}{\sqrt{m_{\kappa}}}\right),
\end{equation}
where $\kappa$ is the index of the atoms, $Z^{*}$ is a Born effective charge, $m_{\kappa}$ is the mass of the $\kappa$-th atom, and $e_{\alpha\kappa}(q)$ is a phonon eigenvector normalized as $\sum_{\kappa\alpha}\left[e_{\kappa\alpha}(\vb*{q}j)\right]^{*}e_{\kappa\alpha}(\vb*{q}j')=\delta_{jj'}$.

The reflectivity $R$ of optical waves normal to the surface is given by
\begin{align}
 R(\omega)=\left|\frac{\sqrt{\epsilon(\omega)}-1}{\sqrt{\epsilon(\omega)}+1}\right|^2. \label{reflectivity}
\end{align}

\subsection{Phonon self-energy}\label{subsec:scph}
\begin{figure}[t]
\centering
\includegraphics[]{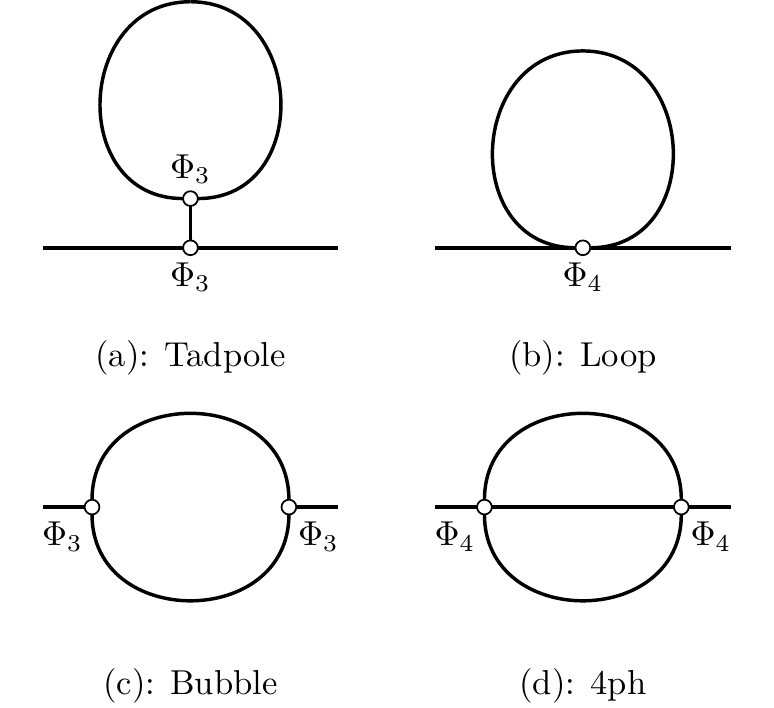}
\caption{Feynman diagrams of phonon self energies. Solid lines and open circles represent phonon propergators and phonon vertexes, respectively.}
\label{fig:diagram}
\end{figure}

Calculating a dielectric function from \cref{143252_19Jul22} requires estimating the phonon self-energy $\Sigma$. As the main contribution to the self-energy, we consider the following terms
\begin{align}
 \Sigma = \Sigma^{\mathrm{T}}+\Sigma^{\mathrm{B}}+\Sigma^{\mathrm{L}}+\Sigma^{\mathrm{4ph}}.
\end{align}
Here, T, B, L, and $4\mathrm{ph}$ stand for tadpole, bubble, loop, and four phonon scattering. \Cref{fig:diagram} depicts the Feynman diagrams of these self-energies. These diagrams are given by the following formulae~\cite{dellavalle1992Equation}.

% \begin{minipage}[b]{0.45\linewidth}
% \centering
% \subfloat[][Tadpole]{\includestandalone{feynman_diagram/tadpole}}
% \end{minipage}
% \begin{minipage}[b]{0.45\linewidth}
% \centering
% \subfloat[][Loop]{\includestandalone{feynman_diagram/loop}}
% \end{minipage}

% \begin{minipage}[b]{0.45\linewidth}
% \centering
% \subfloat[][Bubble]{\includestandalone{feynman_diagram/bubble}}
% \end{minipage}
% \begin{minipage}[b]{0.45\linewidth}
% \centering
% \subfloat[][4ph]{\includestandalone{feynman_diagram/4ph}}
% \end{minipage}

\begin{align}
 \Sigma^{\mathrm{T}}_{q}(\omega)=&\frac{-1}{\hbar}\sum_{q_2,j_1=\mathrm{TO}}V(-q,q,\vb*{0}j_1)V(\vb*{0}j_1,q_2,-q_2)\frac{2n_2+1}{\omega_{\vb*{0}j_1}} \\
 \Sigma^{\mathrm{B}}_{q}(\omega)=&\frac{1}{2\hbar}\sum_{q_1,q_2,s\pm 1}\left|V(-q,q_1,q_2)\right|^2 \nonumber \\
     &\left[\frac{n_1+n_2+1}{s\omega_c+\omega_{q_1}+\omega_{q_2}}-\frac{n_1-n_2}{s\omega_c+\omega_{q_1}-\omega_{q_2}}\right] \label{eq:bubble} \\
 \Sigma^{\mathrm{L}}_{q}(\omega)=&-\sum_{q_1}V(q,-q,q_1,-q_1)\frac{2n_1+1}{2} \\
 \Sigma^{\mathrm{4ph}}_{q}(\omega)=&\frac{1}{6\hbar}\sum_{q_1q_2q_3,s\pm 1}V(-q,q_1,q_2,q_3)V(-q_1,-q_2,-q_3,q) \nonumber\label{eq:4ph}\\
     &\left[\frac{(n_1+1)(n_2+1)(n_3+1)-n_1n_2n_3}{s\omega_c+\omega_{q_1}+\omega_{q_2}+\omega_{q_3}} \right. \nonumber\\
 +& \left. \frac{3n_1(n_2+1)(n_3+1)-(n_1+1)n_2n_3}{s\omega_c-\omega_{q_1}+\omega_{q_2}+\omega_{q_3}}\right] 
\end{align}
Here and in the following, we use $q$ for the shorthand notation of $(\vb*{q},j)$, satisfying $q=(\vb*{q},j)$ and $-q=(-\vb*{q},j)$. $n_i=n(\omega_{q_i})=1/(e^{\beta\hbar\omega_{q_i}}-1)$ is the Bose–Einstein distribution function and $\omega_c=\omega+i0^{+}$ with $0^{+}$ being a positive infinitesimal. In addition, the summation in \cref{eq:bubble} is restricted to the pairs $(\vb*{q}_1,\vb*{q}_2)$ satisfying the momentum conservation $\vb*{q}_1+\vb*{q}_2=\vb*{q}+\vb*{G}$, where $\vb*{G}$ is a reciprocal lattice vector. Similarly, the sum of the 4ph diagram~\eqref{eq:4ph} is limited to the pairs $(\vb*{q}_1,\vb*{q}_2,\vb*{q}_3)$ satisfying $\vb*{q}_1+\vb*{q}_2+\vb*{q}_3=\vb*{q}+\vb*{G}$. $V(q_1,q_2,q_3)$ and $V(q_1,q_2,q_3,q_4)$ are three and four phonon scattering matrices defined as 
\begin{align}
 V(q_1,q_2,q_3)&= \frac{1}{N^{1/2}}\left(\frac{\hbar}{2}\right)^{3/2}\sum_{\substack{\kappa_1\mu_1 \\ l_2\kappa_2\mu_2 \\ l_3\kappa_3\mu_3}}\Phi_{\mu_1\mu_2\mu_3}^{0\kappa_1,l_2\kappa_2,l_3\kappa_3} \nonumber \\
 &\times \frac{e_{\kappa_1 \mu_1}(q_1) e_{\kappa_2\mu_2}(q_2) e_{\kappa_3\mu_3}(q_3)}{\sqrt{m_{\kappa_1}m_{\kappa_2}m_{\kappa_3}}}e^{i(\vb*{q}_2\cdot\vb*{r}_2+\vb*{q}_3\cdot\vb*{r}_3)} ,
\end{align}
\begin{align}
 V(q_1,q_2,q_3,q_4)&= \frac{1}{N}\left(\frac{\hbar}{2}\right)^{2} \sum_{\substack{\kappa_1\mu_1 \\ l_2\kappa_2\mu_2 \\ l_3\kappa_3\mu_3 \\ l_4\kappa_4\mu_4}}\Phi_{\mu_1\mu_2\mu_3\mu_4}^{0\kappa_1,l_2\kappa_2,l_3\kappa_3,l_4\kappa_4} \nonumber \\ 
& \times \frac{e_{\kappa_1\mu_1}(q_1) e_{\kappa_2\mu_2}(q_2)e_{\kappa_3\mu_3}(q_3)e_{\kappa_4\mu_4}(q_4)}{\sqrt{m_{\kappa_1}m_{\kappa_2}m_{\kappa_3}m_{\kappa_4}}}  \nonumber \\
 &\times e^{i(\vb*{q}_2\cdot\vb*{r}_2+\vb*{q}_3\cdot\vb*{r}_3+\vb*{q}_4\cdot\vb*{r}_4)},
\end{align}
where $\mu$ is a Cartesian index, $l$ is the index of unit cells, $\vb*{r}_l$ is the position of the $l$th primitive cell, and $\Phi$ represents the third- and fourth-order IFCs, which is the derivative of the potential energy $U$ with respect to atomic displacements $u$ as follows,
\begin{align} 
 \Phi_{\mu_1\cdots\mu_n}^{l_1\kappa_1,\cdots,l_n\kappa_n}=\frac{\partial U}{\partial u_{\mu_1}(l_1\kappa_1)\cdots \partial u_{\mu_n}(l_n\kappa_n)}.
\end{align}
The tadpole and loop diagrams are real constants, while the bubble and 4ph diagrams are complex numbers that depend on the frequency. Thus, only the bubble and 4ph diagrams contribute to phonon linewidths. We ignore the frequency shifts due to thermal expansion and isotope effect because they are small in rutile \ce{TiO2} at room temperature~\cite{henderson2009Temperature}.

As mentioned before, the anharmonicity of rutile \ce{TiO2} is so strong that these self-energies must be treated in a self-consistent manner. $\Sigma^{\mathrm{T}}$ and $\Sigma^{\mathrm{L}}$ are considered self-consistently in the SCPH theory, and anharmonic phonon frequencies are obtained with solving the following self-consistent equation for $\omega$. 
\begin{align}
\left[ G^{\mathrm{S}}_{q}(\omega) \right]^{-1}= \left[ G^{0}_{q}(\omega) \right]^{-1}-\Sigma^{\mathrm{T}}[G^{\mathrm{S}}]-\Sigma^{\mathrm{L}}[G^{\mathrm{S}}]  \label{eq:scph}
\end{align}
Here, $G^{0}_{q}$ and $G^{\mathrm{S}}_{q}$ are harmonic and SCPH phonon Green's functions, respectively. We write the resultant SCPH frequencies as $\omega^{\mathrm{S}}_{q}$. The SCPH+bubble (SCPH+B) theory~\cite{tadano2022FirstPrinciples} has recently been proposed to consider $\Sigma^{\mathrm{B}}$ in the SCPH theory. After solving the SCPH equation~\eqref{eq:scph}, this method solves the following non-linear equation for $\Omega$, 
\begin{align}
 \Omega^2_{q}=\left(\omega^{\mathrm{S}}_{q}\right)^2-2\omega^{\mathrm{S}}_{q}\Re\Sigma^{\mathrm{B}}_{q}[G^{\mathrm{S}},\Phi_3](\omega=\Omega_q).
\end{align}
In the following, we write the SCPH+B phonon frequencies as $\Omega^{\mathrm{SCPH+B}}$.

After obtaining the anharmonic phonon frequencies and eigenvectors by the SCPH+B equation, the imaginary part of $\Sigma^{\mathrm{B}}$ and $\Sigma^{\mathrm{4ph}}$ are considered in a frequency-dependent form as 
\begin{align}
 \Sigma^{\mathrm{B+4ph}}(\omega) &=\Im \Sigma^{\mathrm{B}}[G^{\rm{SCPH+B}},\Phi_3](\omega) \nonumber\\
&+ \Sigma^{\mathrm{4ph}}[G^{\rm{SCPH+B}},\Phi_4](\omega).
\end{align}
We use not harmonic Green's functions but SCPH+B Green's functions to include the phonon frequencies renormalization effect. We finally obtain a dielectric function by substituting these self-energies into \cref{143252_19Jul22} as 
\begin{align}
 \epsilon_{\alpha\beta}(\omega)&=\epsilon^{\infty}_{\alpha\beta}+ \nonumber  \\
& \frac{1}{v_0}\sum_{j}\frac{S^{j}_{\alpha\beta}}{\left(\Omega^{\mathrm{SCPH+B}}_{\vb*{0}j}\right)^2-\omega^2-2\Omega^{\mathrm{SCPH+B}}_{\vb*{0}j}\Sigma^{\mathrm{B+4ph}}(\omega) }.\label{cowley_final}
\end{align}

All the parameters in \cref{cowley_final} will be determined if second, third, and fourth-order IFCs are provided other than $\epsilon^{\infty}$ and $Z^{*}$, which can be calculated from DFPT. As $\epsilon^{\infty}$ is well known to be overestimated in DFT and the evaluation of $\epsilon^{\infty}$ is outside the scope of our work, we use the experimental values~\cite{devore1951Refractive} of $\epsilon^{\infty}_{xx}=5.91$ and $\epsilon^{\infty}_{zz}=7.20$ in the following calculations.

\section{Results and Discussion}

\subsection{Computational Details}

The IFCs of rutile \ce{TiO2} were calculated from \textit{ab initio} calculations using VASP~\cite{kresse1996Efficient}. The local density approximation (LDA)~\cite{perdew1981Selfinteraction} and the $\mathrm{r}^2$ strongly constrained and appropriately normed ($\mathrm{r}^2$SCAN) meta-GGA~\cite{furness2020Accurate} with the projector augmented-wave method~\cite{kresse1999Ultrasoft} were used for exchange and correlation functionals. The semicore $3s$ and $3p$ states are considered as the valence electrons in the \ce{Ti} pseudopotential. The plane-wave energy and charge cutoffs are $\SI{800}{\electronvolt}$ and $\SI{1200}{\electronvolt}$, respectively. The energy convergence threshold is set at $\SI{1e-9}{\electronvolt}$. 

\begin{figure*}[t]
\captionsetup[subfigure]{font={bf,large}, skip=1pt, margin=-0.7cm,justification=raggedright, singlelinecheck=false}
\centering
\begin{subcaptionblock}{0.6\linewidth}
\subcaption{}
\includegraphics[]{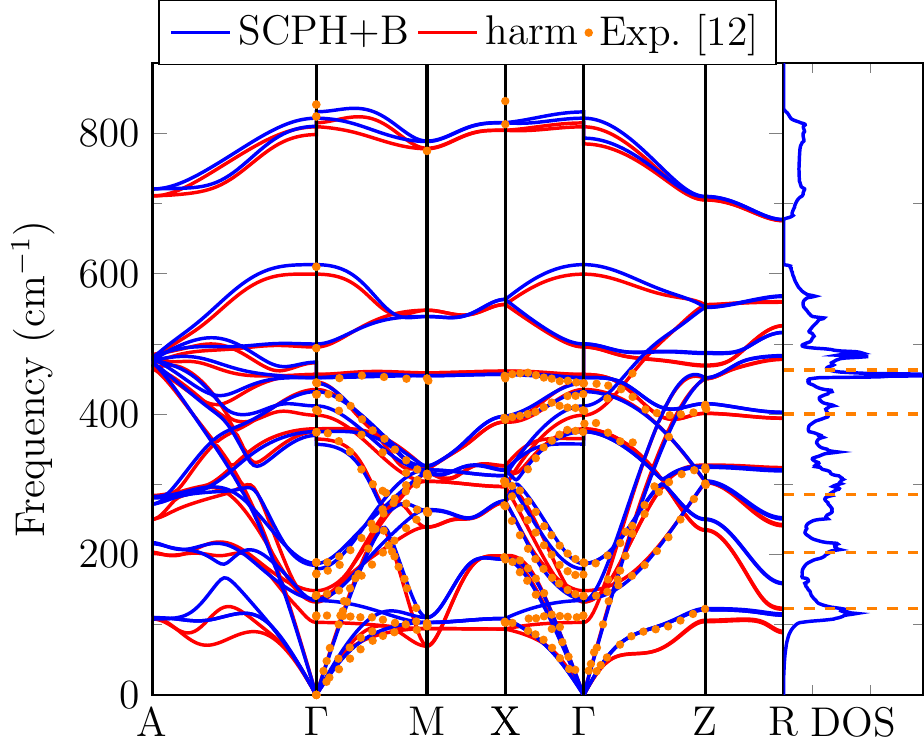}\label{fig:bandr2SCAN}
\end{subcaptionblock}\hfill
\begin{subcaptionblock}{0.4\linewidth}
\subcaption{}
\includegraphics[]{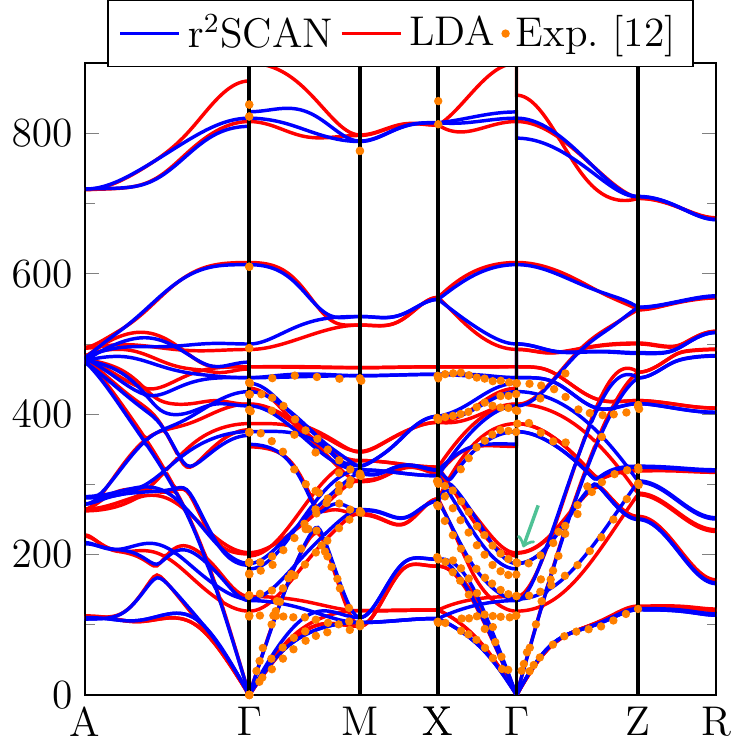}\label{fig:bandLDA}
\end{subcaptionblock}\hfill
  \caption{(a) The harmonic (red) and SCPH+B (blue) band structures of $\mathrm{r}^2$SCAN functional at $300 \mathrm{K}$ with the experimental values from inelastic neutron scattering (orange dot). The SCPH+B DOS is also illustrated at the same time. The orange dashed lines are the peak obtained from the inelastic neutron scattering experiment~\cite{lan2015Phonona}. (b) The SCPH+B band structure with $\mathrm{r}^2$SCAN (red) and LDA (blue) functional. The green arrow indicate the $A_{\mathrm{2u}}$ mode, which is overestimated in the LDA functional.}
  \label{fig:band}
 \end{figure*}
 
Before phonon calculations, the lattice parameter and geometry optimization was performed with the electronic sampling of a $10\times 10\times 10 $ Monkhorst-Pack grid, which converged to the maximum error in forces of $\SI{1}{\milli\electronvolt/\angstrom}$ and stresses of $\SI{0.01}{\giga\pascal}$. Born effective charges are obtained from DFPT calculations for both long-range interactions and dielectric properties.

We estimated IFCs via linear-regression optimization using DFT forces of various atomic configurations as training data. The harmonic terms were fitted from the finite displacement method with one atom moved by $\SI{0.01}{\angstrom}$, where the atomic forces were calculated building a $2\times 2\times 4$ supercell with a $5\times 5\times 5$ electronic wave vector grid. Then, we used the LASSO technique~\cite{zhou2014Lattice} to extract anharmonic IFCs from the displacement-force training datasets with all atoms moved by $\SI{0.04}{\angstrom}$ in random directions.  When constructing the IFC model, we included all possible IFCs in a $2\times 2\times 4$ supercell for harmonic IFCs. The cubic, quartic, fifth, and sixth terms were considered with cutoffs of $15$, $10$, $5$, and $5$ bohr, respectively. 

The SCPH and SCPH+B equations, including $\Sigma^{\mathrm{L}}$ and the real part of $\Sigma^{\mathrm{B}}$, were solved for a $2\times 2\times 2$ $q$ points, where a $6\times 6\times 6$ $q$ points grid was used for computing the self-energies~\cite{tadano2015Selfconsistent}. $\Sigma^{\mathrm{T}}$ is ommited because of it's smallness in rutile \ce{TiO2}. Finally, the imaginary parts of $\Sigma^{\mathrm{B}}$ and $\Sigma^{\mathrm{4ph}}$ are calculated with using a $15\times 15\times 15$ and $10\times 10\times 10$ $q$ points grid, respectively. The extraction of IFCs, the lattice dynamics calculations, and the SCPH calculations were performed using the ALAMODE~\cite{tadano2014Anharmonic} package.

\begin{table}[b] \centering
\caption{calculated lattice constants with LDA and $\mathrm{r}^2$SCAN. Experimental values are taken from a neutron diffraction study~\cite{burdett1987Structuralelectronic}.  $a$ is the lattice constant in the $x$ and $y$ directions, and $c$ in the $z$ direction. $v_0$ is the lattice volume. The four oxygen O ions are located at the $(u,u,0)$, $(1-u,1-u,0)$, $(1/2-u,1/2+u,1/2)$, and $(1/2+u,1/2-u,1/2)$ in the fractional coordinate, where $u$ is a parameter.}

% ----- 表:lattice constants ------
 \renewcommand{\arraystretch}{1.5}
{\tabcolsep = 0.15cm
  \begin{tabular}{lccccc}
   \hline\hline
       & \SI[parse-numbers = false]{a}{(\angstrom)} & \SI[parse-numbers = false]{c}{(\angstrom)} & $u$ & $c/a$ & \SI[parse-numbers = false]{v_0}{(\angstrom^3)} \\
   \hline 
    LDA                    & $4.552$  & $2.922$  & $0.3038$  & $0.642$   & $60.55$ \\
    $\mathrm{r}^2$SCAN     & $4.602$  & $2.961$  & $0.3046$  & $0.643$   & $62.71$ \\
    Exp. \SI{300}{\kelvin}~\cite{burdett1987Structuralelectronic}   & $4.593$ & $2.959$ & $0.3048$ & $0.644$   & $62.42$ \\
%    X ray 298K & $4.5937$ & $2.9587$ & $0.30511$ & $0.644$   & \\ 
    Exp. \SI{15}{\kelvin}~\cite{burdett1987Structuralelectronic}    & $4.587$ & $2.954$ & $0.3047$ & $0.644$   & $62.15$ \\
    \hline
   \hline
  \end{tabular}
}
\renewcommand{\arraystretch}{1.0}

\ifstandalone
\input{../ver_0_after.bbl}
\fi
 \label{table:lattice_constant}
\end{table} 
\begin{table}[t] \centering
\caption{calculated Born effective charge tensors $Z$ of the Ti atom at $(0, 0, 0)$ and the O atom at $(u, u, 0)$. $Z_{xx} = Z_{yy}$, $Z_{yz} = Z_{zy}$, and $Z_{zz}$ are shown considering the symmetry.}

% ----- 表:born effective charges ------
 \renewcommand{\arraystretch}{1.5}
{\tabcolsep = 0.15cm
  \begin{tabular}{lccccccc}
   \hline\hline
       & \multicolumn{3}{c}{Ti} && \multicolumn{3}{c}{O} \\
       & $Z_{xx}$ & $Z_{xy}$ & $Z_{zz}$ && $Z_{xx}$ & $Z_{xy}$ & $Z_{zz}$   \\
   \hline 
    LDA                & $6.34$  & $-1.01$ & $7.66$  && $-3.17$  & $1.81$ & $-3.83$ \\
    $\mathrm{r}^2$SCAN & $5.96$  & $-0.97$ & $7.27$  && $-2.94$  & $1.71$ & $-3.60$ \\ 
  % exp 300K   &  & & $2.9589$ & $0.30476$ & $0.644$ &  & \\ 
  \hline\hline
  \end{tabular}
}
 \renewcommand{\arraystretch}{1.0}

 \label{table:bec}
\end{table} 
 
\subsection{phonon frequencies}\label{subsection:phonon_frequencies}
\begin{table*}[!ht] \centering
\caption{Comparison of the computed mode frequencies (in $\si{\per\cm}$) at the $\Gamma$ point with various experimental data at room temperature. The results from the harmonic approximation (harm), SCPH and SCPH+B are shown for both LDA and $\mathrm{r}^2$SCAN at $\SI{300}{\kelvin}$, while a usual perturbation calculation (non-SC) are shown only for $\mathrm{r}^2$SCAN.}

\begin{threeparttable}
 \centering
 \renewcommand{\arraystretch}{1.5}
{\tabcolsep = 0.15cm
% ----- 表:Γ点振動数 ------
  \begin{tabular}{lcccccccccccc}
   \hline\hline
    & \multicolumn{3}{c}{LDA} & &\multicolumn{4}{c}{$\mathrm{r}^2$SCAN} && neutron~\cite{traylor1971Lattice} & Raman~\cite{porto1967raman}  & FPSQ~\cite{schoche2013Infrared} \\ \cline{2-4}\cline{6-9}
    & harm & SCPH & SCPH+B && harm & SCPH & SCPH+B & non-SC &&  &IR~\cite{eagles1964Polar} &  \\
   \hline 
    \multicolumn{12}{c}{Raman} \\
   $A_{1g}$    &$612.5$ & $620.8$ & $616.6$ && $599.8$ & $613.1$ & $613.3$ & $627.1$ && $610$    & $612$ & -\\ 
   $A_{2g}$    &$395.0$ & $424.0$ & $413.7$ && $456.5$ & $432.9$ & $451.8$ & $458.9$ && NF\tnote{1}       & NF    & -\\ 
   $B_{1g}$    &$134.7$ & $149.6$ & $140.2$ && $146.5$ & $143.0$ & $138.9$ & $139.1$ && $142$    & $143$ & - \\ 
   $B_{2g}$    &$817.5$ & $818.2$ & $817.2$ && $809.6$ & $814.4$ & $821.8$ & $831.4$ && $824$    & $826$ & - \\ 
   $E_g$       &$464.3$ & $476.0$ & $467.5$ && $434.4$ & $463.1$ & $431.5$ & $429.0$ && $445$    & $447$ & - \\ 
    \multicolumn{12}{c}{non-active} \\
   $B^1_{1u}$  &$108.1$ & $121.9$ & $116.6$ && $103.2$ & $141.0$ & $130.0$ & $159.5$ && $113$    &   -   & - \\
   $B^2_{1u}$  &$417.9$ & $414.8$ & $413.8$ && $398.4$ & $420.7$ & $411.7$ & $421.8$ && $406$    &   -   & - \\
    \multicolumn{12}{c}{TO} \\
   $A_{2u} $   &$147.8$ & $227.9$ & $198.0$ && $138.5$ & $210.1$ & $179.4$ & $210.6$ && $172.6$ & $167$ & $172.3$ \\
   $E^1_{u}$   &$149.0$ & $216.8$ & $202.0$ && $132.7$ & $203.6$ & $186.9$ & $238.4$ && $189$   & $183$ & $188.6$ \\ 
   $E^2_{u}$   &$384.3$ & $394.6$ & $386.5$ && $378.8$ & $382.1$ & $374.9$ & $374.3$ && $374$   & $388$ & $379.3$ \\  
   $E^3_{u}$   &$489.0$ & $504.7$ & $492.5$ && $495.7$ & $513.1$ & $500.1$ & $495.0$ && $494$   & $500$ & $500.5$ \\
    \multicolumn{12}{c}{LO} \\
   $A_{2u} $   &$843.9$ & $861.5$ & $854.2$ && $784.9$ & $800.6$ & $793.1$ & $805.2$ && NF      & $811$ & $796.5$ \\  
   $E^1_{u}$   &$354.6$ & $362.7$ & $354.0$ && $364.5$ & $364.9$ & $356.9$ & $360.0$ && $375$   & $373$ & $365.7$ \\ 
   $E^2_{u}$   &$439.1$ & $447.3$ & $436.8$ && $445.5$ & $455.6$ & $443.9$ & $445.7$ && $428$   & $458$ & $444.9$ \\
   $E^3_{u}$   &$882.8$ & $904.9$ & $904.9$ && $815.1$ & $836.0$ & $830.7$ & $851.4$ && $842$   & $806$ & $829.6$ \\  
    \hline
   \hline
  \end{tabular}
}
 \renewcommand{\arraystretch}{1.0}
\begin{tablenotes}
\item[1] NF=not found.
\end{tablenotes}
\end{threeparttable}

\ifstandalone
\input{../ver_0_after.bbl}
\fi

\label{table:gamma}
\end{table*} 
Rutile \ce{TiO2} has a tetragonal unit cell and the $P4_2/mnm$ space group, as shown in \cref{fig:crystal}. Because six atoms are in the unit cell, there are $15$ optical phonon modes and three acoustic phonon modes. The optical phonons at the $\Gamma$ point of the Brillouin zone belong to the following irreducible representations,
\begin{equation}
  \Gamma_{\mathrm{opt}} = A_{\mathrm{1g}} + A_{\mathrm{2g}} + A_{\mathrm{2u}} + 2B_{\mathrm{1u}} + B_{\mathrm{1g}} + B_{\mathrm{2g}} + E_{\mathrm{g}} + 3E_{\mathrm{u}}.
\end{equation}
Expressions with subscript $\mathrm{g}$ are Raman-active, those with $\mathrm{u}$ are infrared-active, while the representations with the $E$ symbol are degenerate. $E_{\mathrm{u}}$ and $A_{\mathrm{2u}}$ are vibrations in the $xy$-plane and $z$-direction, respectively, contributing to the dielectric function's $xy$ and $z$ components. In the $E^{1}_{\mathrm{u}}$ phonon, the softest $E_{\mathrm{u}}$ phonon, and the $A_{\mathrm{2u}}$ phonon, the \ce{Ti} and \ce{O} ions move in opposite directions, whereas in the $E^{2}_{\mathrm{u}}$ and $E^{3}_{\mathrm{u}}$ modes, the two \ce{Ti} ions move in opposite directions, as in \cref{fig:crystal}.

We first present results for optimized lattice constants from LDA and $\mathrm{r}^2$SCAN compared with experimental values at $\SI{15}{\kelvin}$ and $\SI{295}{\kelvin}$ in \cref{table:lattice_constant}.  As in previous studies, LDA slightly underestimates the lattice constants by $0.8\%$. The $\mathrm{r}^2$SCAN functional, a meta-GGA family, shows good agreement with experimental values within $0.4\%$, though the GGA-PBE functional is known to overestimate the lattice constants.

\Cref{table:bec} shows that Born effective charges obtained from $\mathrm{r}^2$SCAN are around $10\%$ smaller than those from LDA. The LDA values agree well with those of previous LDA studies~\cite{lee1994Dielectric,labat2007Density,lee2011Influence, dou2013Comparative}. Based on the LO phonon frequencies results discussed below, Born effective charges calculated from $\mathrm{r}^2$SCAN are considered more accurate than those from LDA.

\begin{figure}[t]
\centering
\includegraphics[]{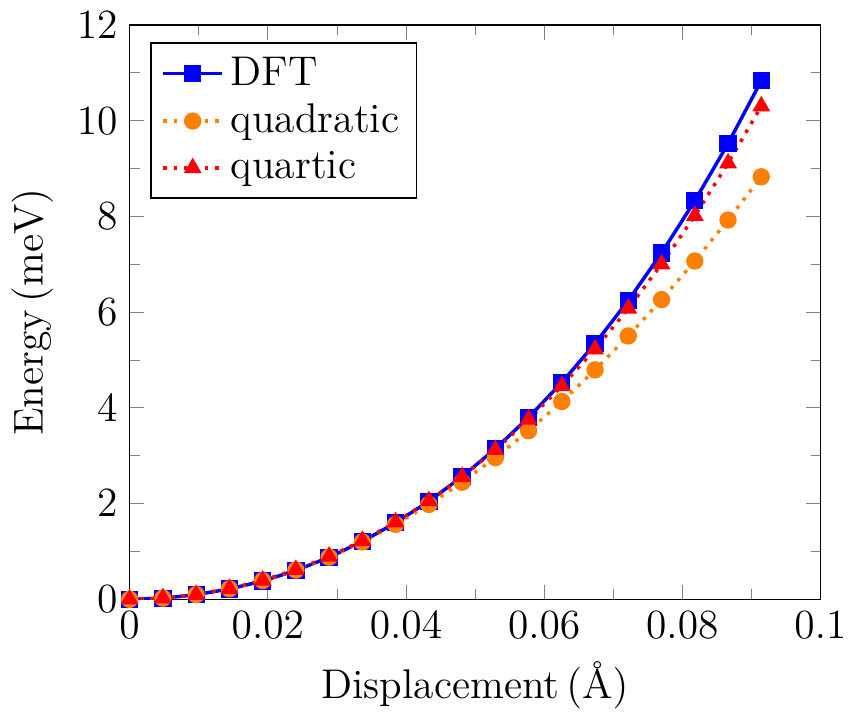} 
\caption{Frozen phonon potential (blue) of $A_{\mathrm{2u}}$ mode with $x$ axis being the displacement of the \ce{Ti} atoms. The quartic component (red) describes the DFT potential well, while the harmonic component (orange) deviates from the potential.}
\label{164402_27Jul22}
\end{figure}

\Cref{fig:band} shows the phonon dispersion spectrum along the high-symmetry points in the first Brillouin zone at $\SI{300}{\kelvin}$ with non-analytic term correction. \Cref{fig:bandr2SCAN} compares harmonic phonon frequencies (red, abbreviated as harm) with SCPH+B frequencies (blue) using the $\mathrm{r}^2$SCAN functional, while \cref{fig:bandLDA} compares the SCPH+B frequencies using LDA (red) and $\mathrm{r}^2$SCAN (blue), together with inelastic neutron scattering results from Traylor~\cite{traylor1971Lattice} (orange dots). \Cref{fig:bandr2SCAN} also shows the density of states (DOS) of $\mathrm{r}^2$SCAN and SCPH+B, with the orange dashed lines being the positions of the five DOS peaks observed in the neutron experiment by Lan and Fultz~\cite{lan2012Phonon}. The combination of the $\mathrm{r}^2$SCAN functional and the SCPH+B calculation agrees well with the experimental data. The frequencies of the $A_{\mathrm{2u}}$ and $E^{1}_\mathrm{u}$ phonons at the $\Gamma$ point and the TA phonon branch, considered highly anharmonic in previous studies, differ significantly between the harmonic approximation and the SCPH+B calculation, with the harmonic approximation predicting smaller frequencies. The potential energy surface of the $A_{\mathrm{2u}}$ phonon is no longer a quadratic function and is well described with considering functions up to the fourth order, as in \cref{164402_27Jul22}. \Cref{table:gamma} summarizes the $\Gamma$ point phonon frequencies. For the $A_{\mathrm{2u}}$ and $E^{1}_{\mathrm{u}}$ modes, the negative frequency shift by the bubble self-energy and the positive frequency shift by the loop self-energy cancel each other out, resulting in about $\SI{50}{\per\cm}$ positive frequency shift. The $A_{\mathrm{2u}}$ phonon frequency within the harmonic approximation is $\SI{139}{\per\cm}$, which rises to $\SI{220}{\per\cm}$ by SCPH. The frequency decreases to $\SI{179}{\per\cm}$ when the bubble self-energy is considered with SCPH+B. The contribution of the anharmonic terms reache $29\%$. Similarly, the $E^{1}_{\mathrm{u}}$ phonon frequency is $\SI{133}{\per\cm}$ for the harmonic approximation, $\SI{210}{\per\cm}$ for SCPH, and $\SI{189}{\per\cm}$ for SCPH+B. Phonon frequencies are also calculated from a usual perturbative approach (abbreviated as non-SC) for $\mathrm{r}^2$SCAN as $\omega=\omega_0+\Delta\omega^{\mathrm{T}}+\Delta\omega^{\mathrm{B}}+\Delta\omega^{\mathrm{L}}$. The non-SC frequencies differ largely from the SCPH+B frequencies in the $A_{\mathrm{2u}}$ and $E^{1}_{\mathrm{u}}$ phonons. In the $A_{\mathrm{2u}}$ phonon mode, we obtained $\Delta\omega^{\mathrm{B}}=-78$ and $\Delta\omega^{\mathrm{L}}=144$, which are too large to be handled within perturbation theory. This calculation shows that neither the harmonic approximation nor the perturbation method suffices for optical properties, where an accurate estimation of the optical phonon frequencies at the $\Gamma$ point is necessary. 

\begin{figure}[t]\centering
\includegraphics[]{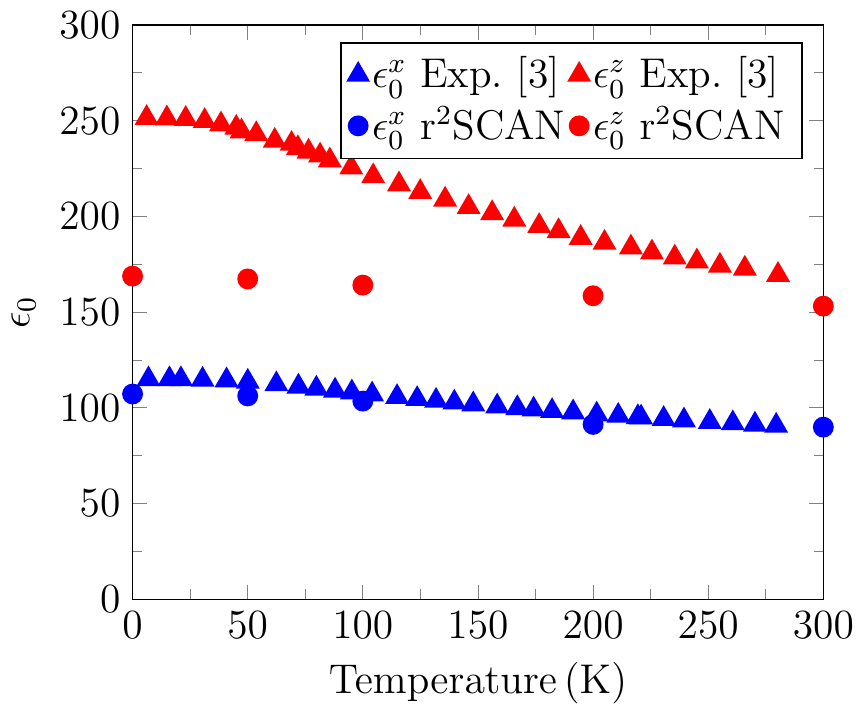}
\caption{Temperature dependence of static dielectric constant. The triangular dots show the experimental values and the circular dots show the results from the SCPH+B calculations.}
\label{134337_5Sep22}
\end{figure} 
 \begin{table}[b] \centering
  \caption{Calculated linewidth ($\si{\per\cm}$) of IR-active phonon modes at $\SI{300}{\kelvin}$ together with the experimental parameters fitted with the FPSQ model. The contributions from the bubble diagram, from the 4ph diagram and the sum of the two are shown for both non-SC and SCPH+B results.}
  \includegraphics{table/lifetime.tex}
  \label{table:lifetime}
 \end{table} 
\begin{figure*}[t]  \captionsetup[subfigure]{font={bf,large}, skip=1pt, margin=-0.7cm,justification=raggedright, singlelinecheck=false}
\centering
\begin{subcaptionblock}{0.48\linewidth}
\subcaption{}\label{fig:dielec}
\adjustbox{right}{\includegraphics[]{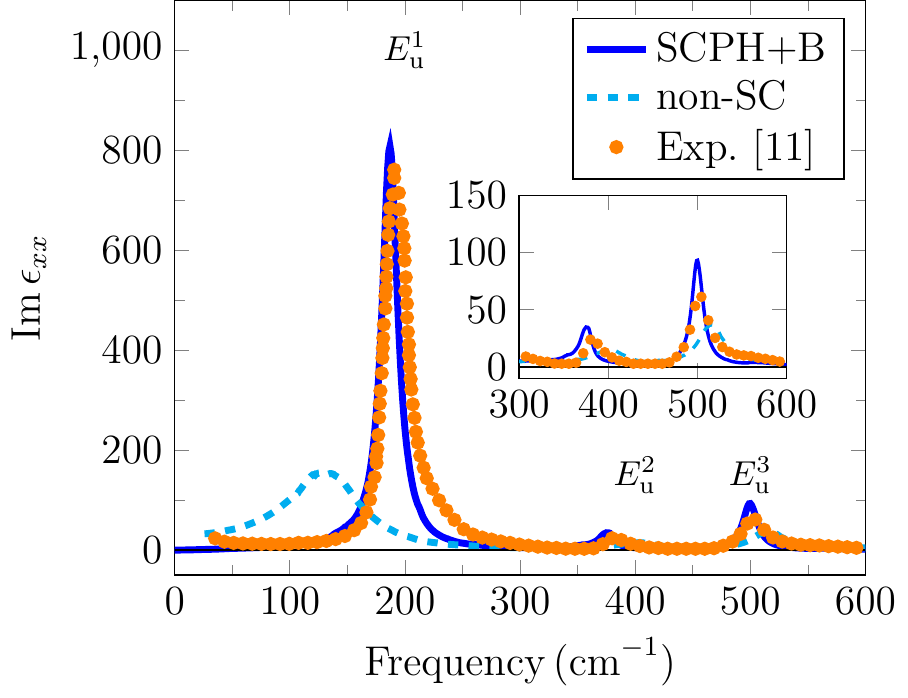}}
\end{subcaptionblock}\hfill
 \begin{subcaptionblock}{0.48\linewidth}
\captionsetup[subfigure]{labelformat=empty}
\adjustbox{right}{\includegraphics[]{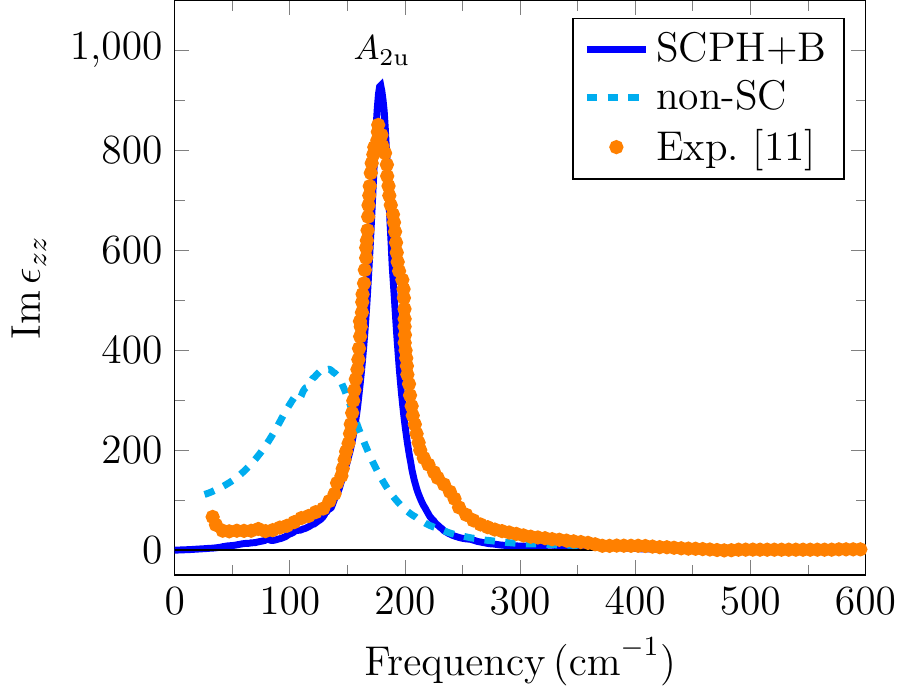}}
\end{subcaptionblock}
\begin{subcaptionblock}{0.48\linewidth}
\subcaption{}\label{fig:reflec}
\adjustbox{right}{\includegraphics[]{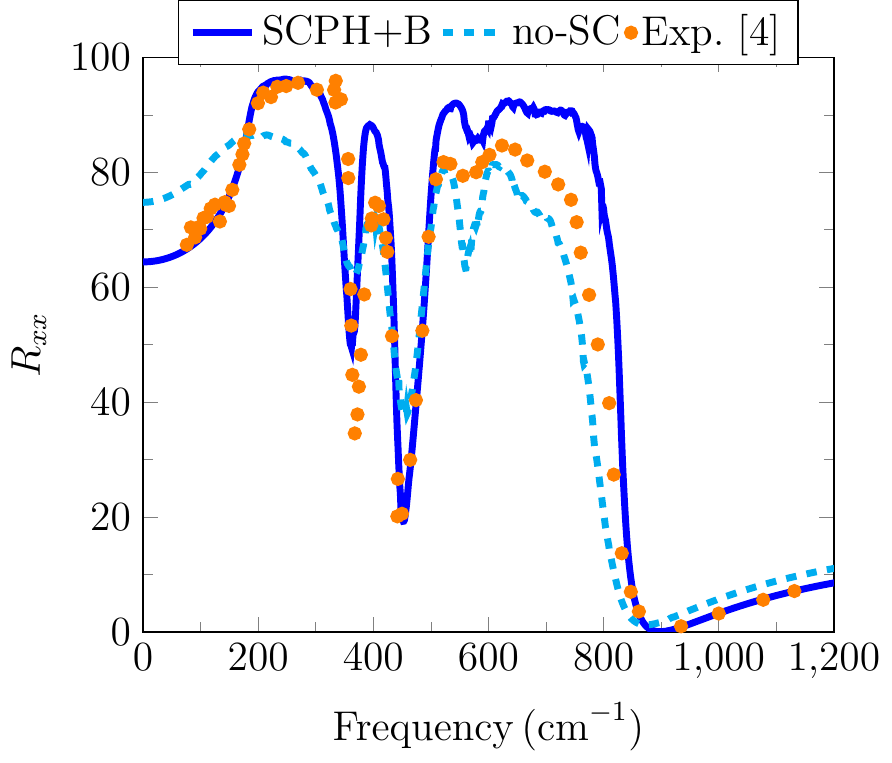}}
\end{subcaptionblock}\hfill
\begin{subcaptionblock}{0.48\linewidth}
\captionsetup[subfigure]{labelformat=empty}
\adjustbox{right}{\includegraphics[]{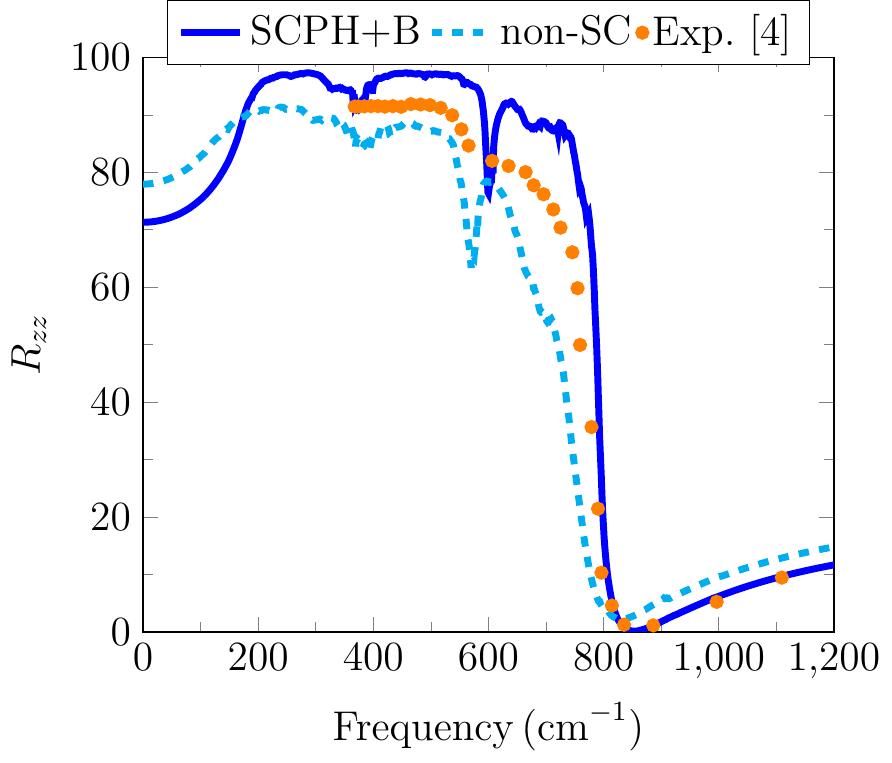}}
\end{subcaptionblock}
\caption{(a) Calculated imaginary part of dielectric functions $\epsilon_{xx}$ (left) and $\epsilon_{zz}$ (right) from SCPH+B (solid blue lines) and harmonic approximation (dashed cyan lines) with experimental data at room temperature~\cite{kanehara2015Terahertz} (orange open circles). The TO phonons corresponding to the peaks are marked. (b) Calculated reflectivity from SCPH+B (solid blue lines) with experimental data at room temperature~\cite{spitzer1962Far}. }
\label{fig:dielreflec}
\end{figure*} 
\Cref{fig:bandLDA} demonstrates a good agreement between between LDA and $\mathrm{r}^2$SCAN throughout the Brillouin zone. However, the LDA calculation overestimates the $A_{\mathrm{2u}}$ phonon at the $\Gamma$ point. The underestimation of the lattice constants of LDA may cause the overestimation of the $A_{\mathrm{2u}}$ phonon, as the $A_{\mathrm{2u}}$ phonon is sensitive to lattice constants~\cite{montanari2002Lattice}. For the LO phonons, the LDA results overestimate the $E^{3}_{\mathrm{u}}$ phonon frequency, the cause of which is larger Born effective charges by LDA than that by $\mathrm{r}^2$SCAN. As $\mathrm{r}^2$SCAN gives better results than LDA, all the following calculations are based on $\mathrm{r}^2$SCAN.

Finally, we calculated the temperature dependence of the static dielectric constant $\epsilon^0\equiv \epsilon(\omega=0)$, which directly reflects the effect of the phonon frequency shift with temperature.  \Cref{134337_5Sep22} compares the calculated temperature dependence of $\epsilon^{0}_x$ and $\epsilon^0_z$ with the experimental data~\cite{parker1961Static}. Both $\epsilon^{0}_x$ and $\epsilon^{0}_z$ increase with lowering temperatures due to a decrease in the phonon frequencies. In particular, $\epsilon^{0}_z$ increases up to $250$ at $T=0$ because of the strong temperature dependence of the $A_{\mathrm{2u}}$ phonon frequency. The SCPH+B calculation well reproduced experimental values for $\epsilon^0_x$. For $\epsilon^{0}_z$, on the other hand, the tendency to increase is reproduced, but the value at $T=0$ is $169$, which is only $70\%$ of the experimental value. 
 
\subsection{phonon linewidth}

\begin{figure*}[t] \centering
\captionsetup[subfigure]{labelformat=empty}
 \subfloat[][]{\includegraphics[]{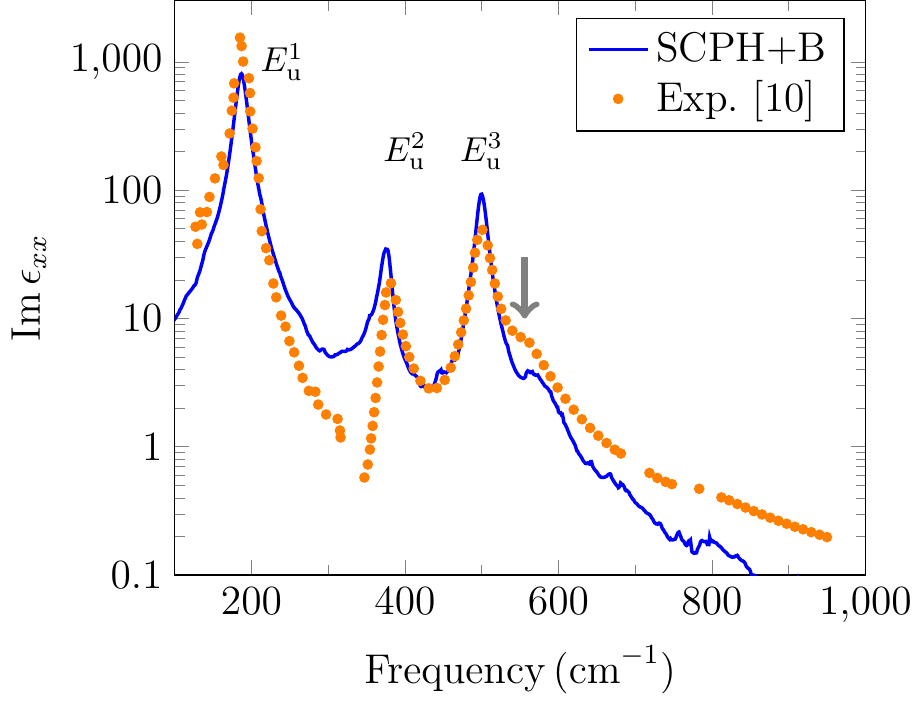}}
 \subfloat[][]{\includegraphics[]{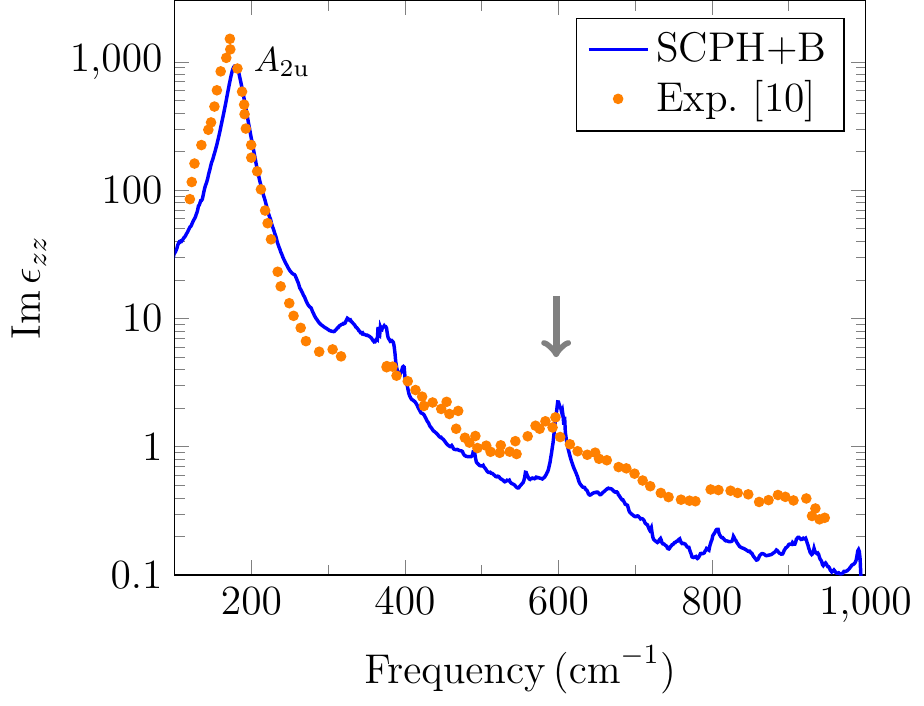}}
\caption{Calculated imaginary part of dielectric functions with experimental data from Schoche~\cite{schoche2013Infrared} in a logarithm scale. Both $\epsilon_{xx}$ and $\epsilon_{zz}$ have a peak at about $\SI{600}{\per\cm}$, which does not belong to any IR-active phonon frequency, and these peaks are indicated by the gray arrows.}
\label{fig:logdiel}
\end{figure*} 
We calculated the frequency-independent linewidths of the four phonon modes involved in the dielectric function in two ways as
\begin{align}
\gamma^{\mathrm{SCPH+B}} &= 2\Im\Sigma[G^{\mathrm{SCPH+B}}](\omega=\Omega^{\mathrm{SCPH+B}}) \\
\gamma^{\mathrm{non-SC}}   &= 2\Im\Sigma[G^{\mathrm{harm}}](\omega=\omega^{\mathrm{harm}}).
\end{align}
As mentioned in \cref{subsec:scph}, while the latter (non-SC) is a usual perturbative calculation, the former uses the SCPH+B phonon frequencies. \Cref{table:lifetime} shows that the Non-SC linewidths are overestimated significantly, whereas the SCPH+B calculations agree better with the FPSQ data~\cite{schoche2013Infrared}, which is determined by fitting experimental reflectivity data to the FPSQ model. It indicates that the calculation of linewidth requires accurate determination of phonon frequencies, including anharmonicity, as pointed out by Fu et al.~\cite{fu2022Finitetemperature}. We also found that self-energies from the four-phonon scattering give a non-negligible contribution in the $A_{\mathrm{2u}}$ and $E^{1}_{\mathrm{u}}$ modes. Such phenomena have been observed in other materials~\cite{yang2020Observationa}.  
\subsection{dielectric function}

\Cref{fig:dielec} shows the calculated imaginary part of dielectric function together with experimental data. The blue line represents the SCPH+B calculation, while the cyan dashed line represents the non-SC calculation, where harmonic phonon frequencies are used, and the frequency-dependent self-energy is calculated as
\begin{align}
 \Sigma^{\mathrm{non-SC}}(\omega) = \Im \Sigma^{\mathrm{B}}[G^{\mathrm{harm}}]+ \Sigma^{\mathrm{4ph}}[G^{\mathrm{harm}}].
\end{align} The maximum values of the imaginary part of $\epsilon_{xx}$ and $\epsilon_{zz}$ reach $785$ and $932$, respectively, which are due to the $E^{1}_{\mathrm{u}}$ and $A_{\mathrm{2u}}$ phonons with large mode-oscillator strength of $S(E^{1}_{\mathrm{u}})=\SI{1.87}{e^2/u}$ and $S(A_{\mathrm{2u}})=\SI{6.12}{e^2/u}$, respectively. It is because the positively charged \ce{Ti} ions and negatively charged \ce{O} ions move in opposite directions in the $A_{\mathrm{2u}}$ and $E^{1}_{\mathrm{u}}$ phonons, as shown in \cref{fig:crystal}. On the other hand, in the $E^{2}_{\mathrm{u}}$ and $E^{3}_{\mathrm{u}}$ phonons, the two \ce{Ti} atoms move in opposite directions, so the mode oscillator strength is much smaller. The SCPH+B calculations agree remarkably well with experimental values, whereas the non-SC calculations failed to reproduce experimental data, especially in the $A_{\mathrm{2u}}$ and $E^{1}_{\mathrm{u}}$ peaks.

\Cref{fig:reflec} shows the reflectivity $R$ in $x$ and $z$ directions calculated using \cref{reflectivity}. For the $x$ direction, the dip due to the $E^{2}_{\mathrm{u}}$ phonon ($\SI{380}{\per\cm}$) is shallower than the experimental data, whereas the dip due to the $E^{1}_{\mathrm{u}}$ and $E^{3}_{\mathrm{u}}$ phonons (bellow $\SI{200}{\per\cm}$ and $\SI{450}{\per\cm}$ ) are in good agreement with experiment. The SCPH+B calculations are overall in better agreement with experiment than the non-SC calculations.

To examine the importance of the frequency dependence of the self-energy, dielectric functions in the logarithm scale are shown in \cref{fig:logdiel}, together with the experimental data from Schoche~\cite{schoche2013Infrared}. The dielectric functions $\epsilon_{xx}$ and $\epsilon_{zz}$ have one peak each at about $\SI{600}{\per\cm}$, which is not the position of any IR-active phonon frequency at the $\Gamma$ point. Several experiments~\cite{schoche2013Infrared, gervais1974Anharmonicity} reported that adding these additional peaks to the FPSQ model improved agreement with experimental data. The peak positions are listed in \cref{fig:additional} together with these experimental data.

\begin{figure}[t]
\centering
  \includegraphics[]{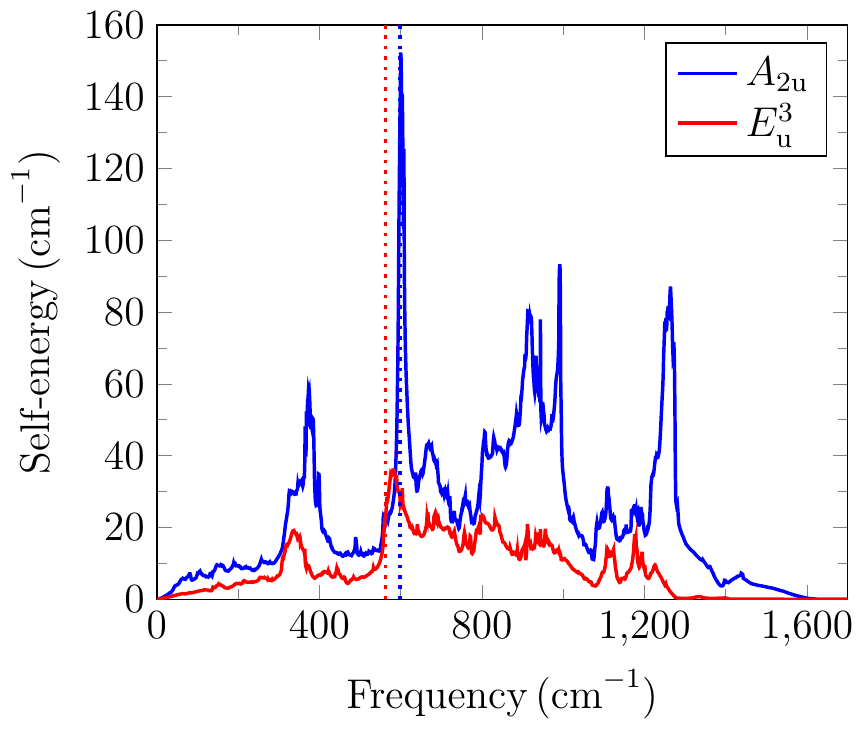}
   \caption{Calculated imaginary part of frequency dependent bubble self-energy of $A_{\mathrm{2u}}$ (blue) and $E^{3}_{\mathrm{u}}$ (red) modes. Blue and red vertical lines represent the additional peaks for $\epsilon_{zz}$ and $\epsilon_{xx}$, respectively.}
   \label{fig:selfenergy}
\end{figure}

\begin{table}[t]
\centering
\caption{Positions of additional peaks (\si{\per\cm}). $\epsilon_{x,1}$ is for $\epsilon_{xx}$ and $\epsilon_{z,1}$ is for $\epsilon_{zz}$.}

% ----- 表:born effective charges ------
\renewcommand{\arraystretch}{1.5}
{\tabcolsep = 0.15cm
  \begin{tabular}{lccc}
   \hline\hline
     Mode     & $\mathrm{r}^2$SCAN & FPSQ~\cite{gervais1974Temperature}    & FPSQ~\cite{schoche2013Infrared}\\
   \hline 
   $\epsilon_{x,1}$ & 563 & $585$ & $556\pm 5$ \\ 
%   $\epsilon_{x,2}$ &     &       & $781\pm 15$ \\ 
   $\epsilon_{z,1}$ & 598 & $592$ & $587\pm 12$ \\ 
%   $\epsilon_{z,1}$ &     &       & $710\pm 18$ \\ 
    \hline
   \hline
  \end{tabular}
}
\renewcommand{\arraystretch}{1.0}

\ifstandalone 
\input{../ver_0_after.bbl}
\fi

\label{fig:additional}
\end{table} 
Possible origins of these peaks, such as lattice defects, have been argued, but the causes are still unclear~\cite{schoche2013Infrared}. The imaginary part of frequency-dependent self-energies can explain these peaks. The frequency dependence of the 4ph self-energy is not so strong, whereas the bubble self-energy has strong frequency dependence in rutile \ce{TiO2}. \Cref{fig:selfenergy} shows the bubble self-energies of the $E^{3}_{\mathrm{u}}$ and $A_{\mathrm{2u}}$ phonons, which contribute to $\epsilon_{xx}$ and $\epsilon_{zz}$, respectively, with vertical dotted lines corresponding to the positions of additional peaks. The bubble self-energies also have peaks at the positions of the additional peaks. We ascribe, therefore, the bubble self-energy to the additional peaks.

 \begin{figure}[t]
 \centering
  \includegraphics[]{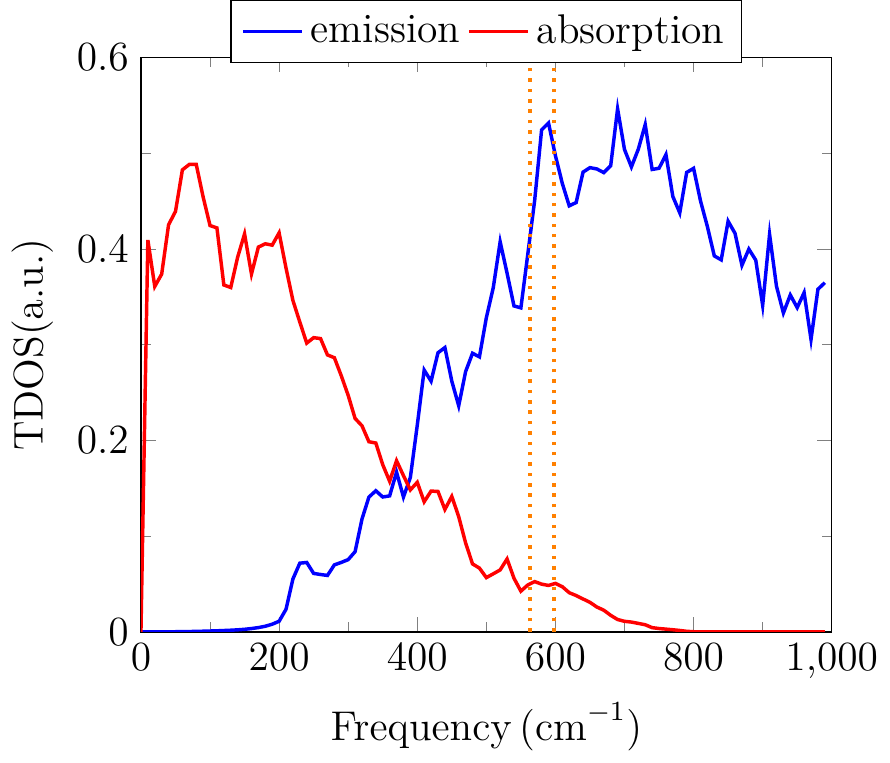}
  \caption{Calculated $\mathrm{TDOS}_{\pm}(\omega,\vb*{q}=\vb*{0})$ for absorption (red) $\omega=\omega_1-\omega_2$ and emission (blue) $\omega=\omega_1+\omega_2$. Red vertical lines represent the positions of additional peaks.}
\label{fig:tdos}
\end{figure} 
When the bubble diagram is taken into account, the dielectric function, and thus the Green's function, has peaks at a certain frequency $\omega$ when the two phonons with frequencies $(\omega_1,\omega_2)$ satisfy the relation $\omega=\omega_1\pm \omega_2$ and $\vb*{q}_1\pm \vb*{q}_2=\vb*{0}$. The positive sign corresponds to phonon emission, and the negative sign to phonon absorption. Such phonon pairs can be specified by the two-phonon density of states (TDOS), which is defined as follows,
\begin{equation}
 \mathrm{TDOS}_{\pm}(\omega,\vb*{q})=\frac{1}{N_q}\sum_{\substack{(\vb*{q}_1,j_1)\\ (\vb*{q}_2,j_2)}}\delta(\omega\pm\omega_{\vb*{q}_1j_1}-\omega_{\vb*{q}_2j_2})\delta_{\vb*{q}\pm \vb*{q}_1,\vb*{q}_2+\vb*{G}}.
\end{equation}
Here $\vb*{G}$ is a reciprocal lattice vector, and $N_q$ is the number of $q$ points in the summation. \Cref{fig:tdos} presents the calculated $\mathrm{TDOS}(\omega,\vb*{q}=\vb*{0})$ with a $15\times 15\times 15$ $q$ points grid, in which TDOS for the emission process has a considerable value at around $\SI{600}{\per\cm}$. Furthermore, from \cref{fig:band}, phonon DOS peaks at around $115$, $210$, $300$, $398$, and $455$ $\si{\per\cm}$, of which the $\SI{115}{\per\cm}$ one is due to acoustic phonons and the others are due to optical phonons. Therefore, it is concluded that the $\SI{600}{\per\cm}$ additional peak is created by the emission process of the pairs $(\SI{115}{\per\cm}, \SI{455}{\per\cm})$, $(\SI{210}{\per\cm}, \SI{398}{\per\cm})$ and $(\SI{300}{\per\cm},\SI{300}{\per\cm})$. Notably, the first pair emission process, involving the acoustic modes, is contributed by phonons with non-zero wave numbers, which can not be detected via single phonon processes by optical probes that are sensitive to $\Gamma$ point phonons. 

\section{Conclusions}

We studied the infrared spectra of rutile \ce{TiO2} using first-principles (DFT) calculations and lattice dynamics calculations. The calculation of phonon frequencies was performed using the SCPH+B theory, a self-consistent anharmonic phonon theory. The SCPH+B calculation very well described the $E^{1}_{\mathrm{u}}$ and $A_{\mathrm{2u}}$ phonon frequencies, which were greatly underestimated in the harmonic approximation. We showed that the anharmonicity in these modes is too strong to treat in a perturbative approach, and self-consistent treatment is essential for accurately describing phonon frequencies. We also compared the LDA and $\mathrm{r}^2$SCAN results, finding that the $\mathrm{r}^2$SCAN functional is more predictive, especially in describing the $A_{\mathrm{2u}}$ mode. 

Phonon linewidths were calculated using both the perturbation theory (non-SC) and the SCPH+B theory. They were significantly overestimated in the non-SC calculation, as suggested by Fu et al.~\cite{fu2022Finitetemperature}. We also found that the contribution from the 4ph self-energy is non-negligible at \SI{300}{\kelvin}. The SCPH+B dielectric function showed good agreement with experimental values. Furthermore, the additional peaks at around $600\mathrm{cm}^{-1}$ pointed out in the previous experiments can be attributed to the two phonon emission process included in the frequency-dependent bubble diagram, which shows the importance of the frequency dependence of phonon self-energies in accurately calculationg the dielectric function. We expect the presented approach to be useful in predicting the dielectric properties of other materials. 

\begin{acknowledgments}
This research was funded by a JST-Mirai Program Grant Number JPMJMI20A1 and a MEXT Quantum Leap Flagship Program (MEXT Q-LEAP) grant number JPMXS0118067246, Japan. T.T. is partially supported by JSPS KAKENHI Grant No. 21K03424. The computations in this study have been done using the facility of the Supercomputer Center, the Institute for Solid State Physics, the University of Tokyo.
\end{acknowledgments}
 
\appendix

\section{Derivation of the Cowley equation}\label{appendix:A}

We review the derivation of the Cowley equation~\eqref{143252_19Jul22}. Consider a supercell with $L$ unit lattices and impose periodic boundary condition. The coordinates of the atoms are denoted by $\vb*{R}$. If the atomic displacements from the equilibrium positions $\vb*{R}_0$ are small compared with the interatomic distance, the dipole moment of the interacting atomic system can be expanded in a power series of the displacements $\vb*{u}(l\kappa) = \vb*{R}(l\kappa)-\vb*{R}_0(l\kappa)$ as 
\begin{align}
 \delta \vb*{M}= \vb*{M}_1+\vb*{M}_2+\vb*{M}_3+\cdots 
\end{align}
where the $\alpha$ component of $\vb*{M}_n$ is
\begin{align}
 M_{n,\alpha}=&\frac{1}{n!}\sum_{l_1\kappa_1\mu_1}\cdots\sum_{l_n\kappa_n\mu_n}M_{\alpha, \mu_1\cdots\mu_n}^{l_1\kappa_1,l_2\kappa_2,\cdots l_n\kappa_n} \nonumber \\
 &\times u_{\mu_1}(l_1\kappa_1)u_{\mu_2}(l_2\kappa_2)\cdots u_{\mu_n}(l_n\kappa_n) \label{eq:Mn}
\end{align}
Here, $\mu$ and $\alpha$ are the indices of Cartesian coordinates, and $u_{\mu}(l\kappa)$ is the displacement of the atom $\kappa$ in the $l$ th cell. The coefficient $M_{\alpha, \mu_1\cdots\mu_n}^{l_1\kappa_1,l_2\kappa_2,\cdots l_n\kappa_n}$ is the $n$th-order derivative of $\vb*{M}$ with respect to atomic coordinates as
\begin{align}M_{\alpha,\mu_1\cdots\mu_n}^{l_1\kappa_1,\cdots,l_n\kappa_n}=\frac{\partial M_{\alpha}}{\partial u_{\mu_1}(l_1\kappa_1)\cdots \partial u_{\mu_n}(l_n\kappa_n)}.
\end{align}
Thus, the first-order coefficient is the Born effective charge as $M_{\alpha,\beta}(l\kappa)=Z^{*}_{\kappa,\alpha\beta}$. From the periodic boundary condition, the value of the quantity does not change when the same number is added to the indices of all cells as
\begin{equation}
 M_{\alpha,\mu_1\cdots\mu_n}^{l_1\kappa_1,\cdots l_n\kappa_n}=
 M_{\alpha,\mu_1\cdots\mu_n}^{0\kappa_1,l_2-l_1\kappa_2,\cdots l_n-l_1\kappa_n} \label{eq:pbc}.
\end{equation}

Next, we introduce the complex normal coordinate $Q_q$, with which the atomic displacement is expressed as \begin{align}
 u_{\mu}(l\kappa)=\frac{1}{\sqrt{Lm_{\kappa}}}\sum_{q}Q_q e_{\mu\kappa}(q)e^{i \vb*{q}\cdot \vb*{r}_l}, \label{qe:normalcord}
\end{align}
By substituting \cref{qe:normalcord} for \cref{eq:Mn} and using \cref{eq:pbc}, we obtain $\vb*{M}_n$ expressed in terms of the normal coordinate as follows,
\begin{align}
 \vb*{M}_n=\frac{1}{n!}\frac{L}{L^{n/2}}\sum_{q_1,\cdots,q_n}\Delta(\vb*{q}_1+\cdots \vb*{q}_n)\vb*{M}(q_1,\cdots,q_n) Q_{q_1}\cdots Q_{q_n},
\end{align}
where
\begin{align}
 \vb*{M}(q_1,\cdots,q_n) =& \sum_{\kappa_1\mu_1}\cdots\sum_{l_n\kappa_n\mu_n}\vb*{M}_{\mu_1\cdots\mu_n}^{0\kappa_1,l_2\kappa_2,\cdots,l_n\kappa_n} \nonumber \\
&\times \frac{1}{\sqrt{m_{\kappa_1}\cdots m_{\kappa_n}}}e_{\mu_1}(q_1,\kappa_1)\cdots e_{\mu_n}(q_n,\kappa_n) \nonumber \\
&\times \exp\left(i(\vb*{q}_2\cdot\vb*{r}_{l_2}+\cdots +\vb*{q}_n\cdot\vb*{r}_{l_n})\right) \label{eq:Mqn}.
\end{align}
$\Delta(\vb*{q})$ takes the value $1$ only when $\vb*{q}$ is the reciprocal lattice vector and $0$ otherwise. Therefore, the summation in first-order expansion is restricted to $\vb*{q}_1=\vb*{0}$, and that of the second-order expansion is restricted to $\vb*{q}_2=-\vb*{q}_1$.

When phonon frequencies of all phonon modes are real in the entire Brillouin zone, one may further transform \cref{eq:Mqn} into a second quantization representation by using $Q_q = (\hbar/2\omega_q)^{1/2}A_q$ with $A_q = b_q + b^{\dagger}_{-q}$ being the displacement operator.
\begin{align}
\vb*{M}_n=&\frac{L}{n!}\left(\frac{\hbar}{2L}\right)^{n/2}\sum_{q_1,\cdots,q_n}\Delta(\vb*{q}_1+\cdots +\vb*{q}_n) \nonumber \\
 &\times \frac{\vb*{M}(q_1,\cdots,q_n)}{\sqrt{\omega_{q_1}\cdots \omega_{q_n}}} A_{q_1}\cdots A_{q_n},\label{eq:quanM}
\end{align}

When an external electric field $\vb*{E}(t)=\vb*{E}_0e^{-i\omega t+\delta t}$ is applied to the system, the interaction is represented by the Hamiltonian as
 \begin{align}
  H_I=-\vb*{M}\cdot \vb*{E}(t).
 \end{align}
According to the linear response theory, the expectation value of the polarization $\vb*{P}=\vb*{M}/v_0$ of the system is
\begin{align}
  \bar{P}_{\alpha}(t)=\frac{1}{v_0} G^{R}(M_{\alpha},\vb*{M},\omega)\cdot \vb{E}(t),
\end{align}
where $G^R(A,B,\omega)$ is the retarded Green's function for operators $A$ and $B$. By using the fact that the polarization and the electric field are connected by the dielectric susceptibility $\vb*{\chi}$ as $P_{\alpha}=\chi_{\alpha\beta}\epsilon_0 E_{\beta}$ and that the dielectric function $\vb*{\epsilon}$ in the IR region is the sum of the phonon contribution $\vb*{\chi}$ and the electron contribution $\vb*{\epsilon}^{\infty}$ as $\epsilon_{\alpha\beta}=\epsilon_{\alpha\beta}^{\infty}+\chi_{\alpha\beta}$, the dielectric function can be written as 
\begin{align}
  \epsilon_{\alpha\beta}(\omega)=\epsilon_{\alpha\beta}^{\infty}+\frac{1}{v_0}G^{R}(M_{\alpha},M_{\beta},\omega).\label{eq:lrt}
\end{align}

\begin{figure}[t]
 \centering
 \includegraphics[]{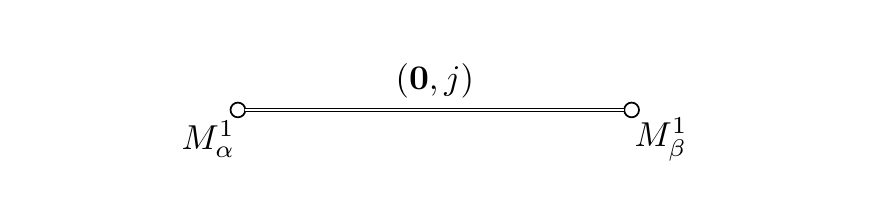}
 \caption{The lowest order of the dielectric function. The double line represents the full-phonon Green's function.}
 \label{fig:diagram_a}
\end{figure}

Substituting \cref{eq:quanM} into \cref{eq:lrt}, the lowest order contribution of the dielectric function, as shown in \cref{fig:diagram_a}, is 
\begin{align}
  \epsilon_{\alpha\beta}(\omega)&=\epsilon_{\alpha\beta}^{\infty}+\frac{1}{v_0}\sum_{jj'}\frac{\hbar}{2}\frac{M(\vb{0}j)M(\vb{0}j')}{\sqrt{\omega_{\vb{0}j}\omega_{\vb{0}j'}}}G^{R}(A_{\vb{0}j},A_{\vb{0}j'},\omega) \nonumber \\
&=\epsilon^{\infty}_{\alpha\beta}+\frac{1}{v_0}\sum_{(\vb*{0},j)}\frac{S^{j}_{\alpha\beta}}{\left(\omega_{\vb*{0}j}\right)^2-\omega^2-2\omega_{\vb*{0}j} \Sigma_{\vb*{0}j}(\omega)}.
\end{align}
 
% \bibliographystyle{apsrev4-2}
% \bibliography{../references/tio2_ref.bib}


\end{document}